\documentclass[twoside,9pt]{extarticle}

\usepackage{lipsum} 

\usepackage[sc]{mathpazo} 
\usepackage[T1]{fontenc} 
\usepackage{microtype} 

\usepackage[hmarginratio=1:1,top=32mm,columnsep=20pt]{geometry} 
\usepackage{multicol} 
\usepackage[hang, small,labelfont=bf,up,textfont=it,up]{caption} 
\usepackage{booktabs} 
\usepackage{float} 
\usepackage[hidelinks]{hyperref}
\usepackage{float}

\usepackage{lettrine} 
\usepackage{indentfirst}
\usepackage{paralist} 
\usepackage{abstract} 

\usepackage{titlesec} 
\renewcommand\thesection{\Roman{section}} 
\renewcommand\thesubsection{\Alph{subsection}} 
\titleformat{\section}[block]{\LARGE\scshape\centering}{\thesection.}{1em}{} 
\titleformat{\subsection}[block]{\large}{\thesubsection.}{1em}{} 

\usepackage{fancyhdr} 
\usepackage{pbox} 
\newcommand*{\nkeywfont}{\fontfamily{pcr}\selectfont}

\newcommand*{\pbrow}[1]{
  \pbox[t]{\linewidth}{\nkeywfont #1\unskip\strut} 
}

\usepackage{graphicx}
\usepackage{epstopdf}

\usepackage[square, numbers]{natbib}
\providecommand{\keywords}[1]{\textbf{\textit{Keywords: }} #1}

\title{\vspace{-14mm}\fontsize{16pt}{10pt}\selectfont\textbf{Efficient fourth order symplectic integrators\\ for near-harmonic separable Hamiltonian systems}} 
\author{
\large
\textsc{Kristian Mads Egeris Nielsen}\thanks{\href{mailto:KristianEgeris@zymplectic.com}{KristianEgeris@Zymplectic.com}}
}
\date{}
\begin{document}
\maketitle
\thispagestyle{fancy}
\begin{abstract}
Efficient fourth order symplectic integrators are proposed for numerical integration of separable Hamiltonian systems $H(p,q)=T(p)+V(q)$. Symmetric splitting coefficients with five to nine stages are obtained by higher order decomposition of the simple harmonic oscillator. The performance of the methods is evaluated for various Hamiltonian systems: Integration errors are compared to those of acclaimed integrators composed by S. Blanes et al. (2013), W. Kahan et al. (1999) and H. Yoshida (1990). Numerical tests indicate that the integrators obtained in this paper perform significantly better than previous integrators for common Hamiltonian systems.\\ 
\end{abstract}
\begin{multicols}{2}
\section{Introduction}
Explicit symplectic integrators (SIs, singular SI) are numerical integration schemes for separable Hamiltonian systems. SIs are widely used in fields of celestial mechanics, accelerator physics, molecular dynamics, and quantum chemistry due to their structure-preserving properties and simple implementation.\\

Numerous SIs with increased accuracy and stability have been published since the first discovery of the fourth order SI \cite{RU}. The perhaps most acclaimed splitting composition is derived by H. Yoshida who proved that SIs of arbitrarily high even order exist and can be constructed using Lie algebra \cite{YO}. Other more recent studies on the topic of high stability SIs include force gradient schemes \cite{CH1} and symplectic correctors \cite{WI}\cite{BL}. The integration order is often associated with the efficiency of numerical integration schemes. $N$th order integrators have their integration error reduced by a factor $k^N$ when the time step is reduced by a factor $k$, making high order integrators highly favorable for small time steps. However, higher order error terms cannot be neglected for large time steps. Symplectic correctors have been proposed for eliminating high order errors of SIs, although these methods are only competitive for specific systems. The force-gradient composition has been proposed as a more efficient splitting scheme for general separable Hamiltonian systems but requires evaluation of the force gradient \cite{CH1}. The force-gradient compositions benefit from all-positive splitting coefficients meaning that only forward steps are taken in phase space and are therefore referred to as a \textit{forward} SIs. Non-gradient splitting methods of orders higher than two always involve negative coefficients \cite{BL}. To construct more efficient high order SIs it has been proposed that the absolute value of negative coefficients and the sum of positive coefficients should be minimized \cite{BL}.\\

This paper presents a series of efficient fourth order SIs which are ideal for numerical integration of particle systems with quadratic kinetic energy. Additionally, some of the presented integrators exhibit higher order accuracy for systems that resemble the harmonic oscillator. The presented splitting coefficients generally have small absolute values and are therefore referred to as \textit{near-forward} SIs.

\section{Fundamental theorems}

Consider the general separable Hamiltonian $H(q,p)=V(q)+T(p)$ with $\dot p =  -\partial H/ \partial q$ and $\dot q = \partial H/ \partial p$ where $q$ and $p$ are canonical coordinates conventionally labeled position and momentum respectively. By defining $z=(q,p)$, the time evolution of $z$ can be expressed by \cite{YO}: 
\begin{equation} \label{eq:1}
z(\tau) = e^{\tau (A+B)}z(0)\
\end{equation}
where $A$ and $B$ are non-commutative operators associated with $T$ and $V$, and $\tau$ is the time step. A set of real numbers exist $(c_1,c_2,...,c_k)$ and $(d_1,d_2,...,d_k)$, referred to as \textit{splitting coefficients}, that satisfy $\sum\nolimits_{i = 1}^k {{c_i}}  = \sum\nolimits_{i = 1}^k {{d_i}}  = 1$ such that the time evolution of $z$ can be approximated by a product of exponential functions \cite{YO}:
\begin{equation} \label{eq:2}
{e^{\tau \left( {A + B} \right)}} = \prod\limits_{i = 0}^k {{e^{{c_i}\tau A}}{e^{{d_i}\tau B}}}  + o\left( {{\tau ^{N + 1}}} \right)\
\end{equation}
where $o(\tau^{N+1} )$ denotes an error on the $N+1$ order term. It is common practice to expand the exponential terms by power series with respect to $\tau$. Lie algebra can be applied to determine the splitting coefficients which satisfy Eq. \ref{eq:2} exactly up to a given order.
The time evolution of $z\left( 0 \right) \to z\left( \tau  \right)$ or $\left( {{q_0},{p_0}} \right) \to \left( {{q_k},{p_k}} \right)$ is explicitly computable as \cite{YO}:
\begin{equation} \label{eq:3}
\begin{array}{l}
{q_i} = {q_{i - 1}} + \tau {c_i}\frac{{\partial T}}{{\partial p}}\left( {{p_{i - 1}}} \right)\\
{p_i} = {p_{i - 1}} - \tau {d_i}\frac{{\partial V}}{{\partial q}}\left( {{q_i}} \right)
\end{array}\
\end{equation}
for $i = 1,2,...,k$. Note that the intermediate steps $(q_j,p_j )$,$j=1,2,...,k-1$, referred to as \textit{substeps}, do not represent the time evolution of $(q_0,p_0)$.
SIs are time reversible if the symplectic coefficients are symmetric \cite{YO}. 
Time reversibility implies that $z\left( 0 \right) \to z\left( \tau  \right)$ can be exactly reverted by $z\left( 0 \right) \leftarrow z\left( { - \tau } \right)$. This property is embraced throughout this paper. The symmetric composition in the form of Eq. \ref{eq:3} with $d_k=0$ is referred to as \textit{ABA} \cite{BL}. If the roles of $A$ and $B$ in Eq. \ref{eq:2} are switched the time evolution is explicitly computable as: 
\begin{equation} \label{eq:4}
\begin{array}{l}
{p_i} = {p_{i - 1}} - \tau {d_i}\frac{{\partial V}}{{\partial q}}\left( {{q_{i - 1}}} \right)\\
{q_i} = {q_{i - 1}} + \tau {c_i}\frac{{\partial T}}{{\partial p}}\left( {{p_i}} \right)
\end{array}\
\end{equation}
for $i = 1,2,...,k$. The symmetric composition in the form of Eq. \ref{eq:4} with $c_k = 0$ is referred to as \textit{BAB}. Table \ref{table:1} summarizes the symmetric structures of the \textit{ABA}- and \textit{BAB}-composition. Note that the $c$ and $d$ coefficients are consistently used in connection with the $\partial T/ \partial p$ and $\partial V/ \partial q$ term respectively. The computational cost of SIs is approximately proportional to the number of derivative evaluations (\textit{stages}). The number of stages is defined as $s=k-1$ for symmetric coefficients due to the \textit{first same as last} property, which implies that the first derivative evaluation is the same as the last derivative evaluation of the previous iteration.
\begin{table}[H] 
\caption{\textit{ABA} and \textit{BAB} composition for symmetric coefficients}
\label{table:1}
\centering
\begin{tabular}{ll|ll}
\toprule
\multicolumn{2}{c}{\textit{ABA}} & \multicolumn{2}{c}{\textit{BAB}} \\
\cmidrule(r){1-4}
$d_1=d_{k-1}$ & $c_1=c_k$ & $d_1 = d_k$ & $c_1 = c_{k-1}$ \\
$d_2=d_{k-2}$ & $c_2=c_{k-1}$ & $d_2 = d_{k-1}$ & $c_2 = c_{k-2}$ \\
\vdots & \vdots & \vdots & \vdots \\
$d_{k-2}$ &  &  & $c_{k-2}$ \\
$d_{k-1}$ & $c_{k-1}$ & $d_{k-1}$ & $c_{k-1}$ \\
$d_k = 0$ & $c_k$ & $d_k$ & $c_k = 0$ \\
\bottomrule
\end{tabular}
\end{table}
\section{Methods}
This section first describes a decomposition method which follows directly from the composition of the simple harmonic oscillator. It is then shown that this decomposition can be used to satisfy general fourth order conditions and higher order conditions for the simple harmonic oscillator.\\

A single iteration from $(q_a,p_a)$ to $(q_b,p_b)$ using the $s$-stage \textit{BAB}-composition (Eq. \ref{eq:4}) can be expanded into:
\begin{equation} \label{eq:5}
\begin{array}{l}
{p_1} = {p_a} - \tau {d_1}\frac{{\partial V}}{{\partial q}}\left( {{q_a}} \right)\\
{q_1} = {q_a} + \tau {c_1}\frac{{\partial T}}{{\partial p}}\left( {{p_1}} \right)\\
{p_2} = {p_1} - \tau {d_2}\frac{{\partial V}}{{\partial q}}\left( {{q_1}} \right)\\
 \vdots \\
{q_s} = {q_{s - 1}} + \tau {c_s}\frac{{\partial T}}{{\partial p}}\left( {{p_s}} \right) = {q_b}\\
{p_{s + 1}} = {p_s} - \tau {d_{s + 1}}\frac{{\partial V}}{{\partial q}}\left( {{q_s}} \right) = {p_b}
\end{array}\
\end{equation}
Suppose now that ${{\partial T\left( p \right)} \mathord{\left/
 {\vphantom {{\partial T\left( p \right)} {\partial p}}} \right.
 \kern-\nulldelimiterspace} {\partial p}} = {p \mathord{\left/
 {\vphantom {p m}} \right.
 \kern-\nulldelimiterspace} m}$
which applies to many classical systems including the simple harmonic oscillator. Eq. \ref{eq:5} then yields:
\begin{equation} \label{eq:6}
\begin{array}{l}
{q_1} = {q_a} + \tau {c_1}\frac{{{p_a}}}{m} + {\tau ^2}{D_1}{m^{ - 1}}\frac{{\partial V}}{{\partial q}}\left( {{q_a}} \right)\\
{q_2} = {q_1}\left( {1 + {C_2}} \right) - {C_2}{q_a} + {\tau ^2}{D_2}{m^{ - 1}}\frac{{\partial V}}{{\partial q}}\left( {{q_1}} \right)\\
{q_3} = {q_2}\left( {1 + {C_3}} \right) - {C_3}{q_1} + {\tau ^2}{D_3}{m^{ - 1}}\frac{{\partial V}}{{\partial q}}\left( {{q_2}} \right)\\
 \vdots \\
{q_s} = {q_{s - 1}}\left( {1 + {C_s}} \right) - {C_s}{q_{s - 2}} + {\tau ^2}{D_s}{m^{ - 1}}\frac{{\partial V}}{{\partial q}}\left( {{q_{s - 1}}} \right)\\
{p_{s + 1}} = m\frac{{{q_s} - {q_{s - 1}}}}{{\tau {c_s}}} - \tau {d_{s + 1}}\frac{{\partial V}}{{\partial q}}\left( {{q_s}} \right)
\end{array}\
\end{equation}
where $C_n=c_{n+1}/c_n$ and $D_n=-c_n d_n$. The \textit{first same as last} property allows substitution of $p_a$ by $p_{s+1}$ yielding an expression that is independent of $p$. Eq. \ref{eq:6} yields significantly larger truncation errors than Eq. \ref{eq:5} and is therefore not suitable for practical implementation. However, Eq. \ref{eq:6} allows separation of ${\tau ^\lambda }$ terms for any order $\lambda=0,1,...,\lambda_{max}$ where $\lambda_{max}$ is the maximum number of error terms which appear by decomposition. For instance consider the three stage decomposition expressed by the entries ${Q_{h,\lambda }}$ in Table \ref{table:2} where $h$ denotes the intermediate step number. This decomposition follows directly from Eq. \ref{eq:6}. A similar decomposition can be constructed for the \textit{ABA} composition. The position contribution $\zeta_{\lambda}$ for each order $\lambda$ is introduced as:
\begin{equation} \label{eq:7}
{\zeta _\lambda }{\tau ^\lambda } \equiv {Q_{s,\lambda }}{\tau ^\lambda },\lambda  = 0,1,...,{\lambda _{\max }}\
\end{equation}
where $s$ is the number of stages. 
The new position can then be calculated as the sum of all position contributions:
\begin{equation} \label{eq:8}
{q_b} = \sum\limits_{\lambda  = 0}^{{\lambda _{\max }}} {{\tau ^\lambda }{\zeta _\lambda }}  + o\left( {{\tau ^{{\lambda _{\max }} + 1}}} \right)\
\end{equation}
\begin{table*}[t] 
\begin{center}
\begin{changemargin}{-0.20cm}{0.0cm} 
\caption{\textit{BAB} decomposition for symmetric coefficients where $\bar C_h = C_h + 1$, ${\bar Q_{h,\lambda }}=m^{-1}\partial V/\partial q(Q_{h,\lambda})$ and $Q_{h,\lambda}$ denotes the table value at indices $(h,\lambda)$}
\label{table:2}
\begin{tabular}{l|l|l|l|l|l|l|l}
\toprule
${h^{{\tau ^\lambda }}}$ & $\tau^0$ & $\tau^1$ & $\tau^2$  & $\tau^3$ & $\tau^4$ & $\tau^5$ & $\tau^6$ \\
\cmidrule(r){1-8}
$0$ & $q_a$ & & & & & & \\
$1$ & $Q_{0,0}$ & $c_1p_am^{-1}$ & $D_{1}\bar Q _{0,0}$  & &  & & \\
$2$ & $Q_{1,0}\bar C_2-C_2Q_{0,0}$ & $Q_{1,1}\bar C_2$  & $Q_{1,2}\bar C_2+D_2\bar Q_{1,0}$  & $D_2\bar Q_{1,1}$ & $D_2\bar Q_{1,2}$  & & \\
$3$ & $Q_{2,0}\bar C_3-C_3Q_{1,0}$ & $Q_{2,1}\bar C_3 - C_3Q_{1,1}$ & $Q_{2,2}\bar C_3-C_3Q_{1,2} + D_3\bar Q_{2,0}$  & $Q_{2,3}\bar C_3 + D_3\bar Q_{2,1}$ & $Q_{2,4}\bar C_3+D_3\bar Q_{2,2}$ & $D_3\bar Q_{2,3}$ & $D_3\bar Q_{2,4}$\\
\bottomrule
\end{tabular}
\end{changemargin}
\end{center}
\end{table*}
Suppose that a known function $q(t)$ exists such that $q(t_0)=q_a$ and $q(t_0+\tau)=q_b+o(\tau^{\lambda_{max}+1} )$ then $q_b$ is expressed by the power series:
\begin{equation} \label{eq:9}
\begin{array}{l}
q\left( {{t_0} + \tau } \right)\\ = {q_a} + \frac{{q'\left( {{t_0}} \right)}}{{1!}}{\tau ^1} + \frac{{q''\left( {{t_0}} \right)}}{{2!}}{\tau ^2}+ ... + \frac{{{q^{\left( {{\lambda _{\max }}} \right)}}\left( {{t_0}} \right)}}{{{\lambda _{\max }}!}}{\tau ^{{\lambda _{\max }}}} + o\left( {{\tau ^{{\lambda _{\max }} + 1}}} \right)\\
 = {\zeta _0} + {\zeta _1}{\tau ^1} + {\zeta _2}{\tau ^2} + ... + {\zeta _{{\lambda _{\max }}}}{\tau ^{{\lambda _{\max }}}} + o\left( {{\tau ^{{\lambda _{\max }} + 1}}} \right)
\end{array}\
\end{equation}
If the SI is of $N$th order then
\begin{equation} \label{eq:10}
\frac{{{q^{\left( \lambda  \right)}}\left( {{t_0}} \right)}}{{\lambda !}}{\tau ^\lambda } = {\zeta _\lambda }{\tau ^\lambda },\lambda  = 0,1,...,N\
\end{equation}
for any function $q(t)$ that qualifies as a time evolution in $H(p,q)=T(p)+V(q)$.
The simple harmonic oscillator (SHO): 
\begin{equation} \label{eq:11}
H = \frac{{{p^2}}}{{2m}} + \frac{1}{2}k{q^2}\
\end{equation}
is enforced by applying the condition $\bar Q_{h,\lambda} = \partial V/\partial q(Q_{h,\lambda})$ in Table \ref{table:2}. The SHO yields the well-known analytical time evolution of $q$:
\begin{equation} \label{eq:12}
q\left( t \right) = {a}\cos \left( {\sqrt {\frac{k}{m}} t} \right) + {b}\sin \left( {\sqrt {\frac{k}{m}} t} \right)\
\end{equation}
Using Eq. \ref{eq:10} where $\zeta_0,\zeta_1,...,\zeta_6$ are evaluated in Table \ref{table:2} the following identities must be true for symmetric fourth order splitting coefficients:
\begin{equation} \label{eq:13}
\begin{array}{l}
{h_1} = {\zeta _0} = {q_a}\\
\frac{1}{{1!}}\sqrt {\frac{k}{m}} {b} = {\zeta _1} = \frac{{{p_a}}}{m}\left( {2{c_1} + {c_2}} \right)\\
 - \frac{1}{{2!}}\frac{k}{m}{a} = {\zeta _2} =  - \frac{{k{q_a}}}{m}\left( {{d_1} + {d_2}} \right)\left( {2{c_1} + {c_2}} \right)\\
 - \frac{1}{{3!}}{\left( {\frac{k}{m}} \right)^{\frac{3}{2}}}{b} = {\zeta _3} =  - 2{c_1}{d_2}\frac{{k{p_a}}}{{{m^2}}}\left( {{c_1} + {c_2}} \right)\\
\frac{1}{{4!}}{\left( {\frac{k}{m}} \right)^2}{a} = {\zeta _4}= {c_1}{d_2}\frac{{{k^2}{q_a}}}{{{m^2}}}\left( {2{c_1}{d_1} + 2{c_2}{d_1} + {c_2}{d_2}} \right)
\end{array}\
\end{equation}
Elimination of the constants a and b yields the identities:
\begin{equation} \label{eq:14}
\begin{array}{l}
1=2(d_1+d_2)(2c_1+c_2)\\
1=12c_1 d_2 (c_1+c_2)/(2c_1+c_2)\\
1=24c_1d_2(2c_1d_1+2c_2d_1+c_2d_2)
\end{array}\
\end{equation}
It is readily found that the only real solution satisfying Eq. \ref{eq:14} is the one originally found by \cite{RU}:
\begin{equation} \label{eq:15}
\begin{array}{l}
{d_1} = {d_4} = \frac{1}{{2\left( {2 - {2^{1/3}}} \right)}} = \frac{1}{2} - {d_2} = \frac{1}{2} - {d_3}\\
{c_1} = {c_3} = \frac{1}{{2 - {2^{1/3}}}} = \frac{1}{2}\left( {1 - {c_2}} \right)\\
\end{array}\
\end{equation}
A closer examination of Table \ref{table:2} reveals that $m$, $k$ and $q_a$ only appear as scalars of $\zeta_0,\zeta_1,...,\zeta_{\lambda_{max}}$ the $\tau^0$, and that $\tau^2$ and $\tau^4$ separations can be expressed as in Table \ref{table:3}.

\begin{table}[H]
\caption{Decomposition of even orders expressed in terms of the splitting coefficients. $m$ and $q_a$ are set to 1}
\label{table:3}
\centering
\begin{tabular}{l|l|l|l}
\toprule
${h^{{\tau ^\lambda }}}$ & $\tau^0$ & $\tau^2$ & $\tau^4$  \\
\cmidrule(r){1-4}
$0$ & $1$ & & \\
$1$ & $1$ & $- \sum\limits_{i = 1}^1 {{c_i}} \sum\limits_{j = 1}^1 {{d_j}}$ & \\
$2$ & $1$ & $ -\sum\limits_{i = 1}^2 {{c_i}\sum\limits_{j = 1}^i {{d_j}} }$ & $ - \sum\limits_{m = 1}^1 {{Q_{m,2}}{d_{m + 1}}\sum\limits_{k = m + 1}^1 {{c_k}} } $ \\
$\vdots$ & $\vdots$ & $\vdots$ & $\vdots$ \\
$h$ & $1$ & $ -\sum\limits_{i = 1}^h {{c_i}\sum\limits_{j = 1}^i {{d_j}} }$ & $-\sum\limits_{m = 1}^{h - 1} {{Q_{m,2}}{d_{m + 1}}\sum\limits_{k = m + 1}^{h - 1} {{c_k}} }$  \\
\bottomrule
\end{tabular}
\end{table}
Symmetric coefficients omit odd order error terms \cite{YO}, implying that the following identities must be true to satisfy the fourth conditions for the SHO:  
\begin{equation} \label{eq:16}
\begin{array}{l}
{\zeta _2} = \frac{{q''\left( 0 \right)}}{{2!}} =  - \frac{1}{{2!}} =  - \sum\limits_{i = 1}^h {{c_i}\sum\limits_{j = 1}^i {{d_j}} } \\
{\zeta _4} = \frac{{q''''\left( 0 \right)}}{{4!}} = \frac{1}{{4!}} =  - \sum\limits_{m = 1}^{h - 1} {{Q_{m,2}}{d_{m + 1}}\sum\limits_{k = m + 1}^h {{c_k}} } 
\end{array}\
\end{equation}
It has previously been shown that Eq. \ref{eq:16} is in fact general second and fourth order criteria \cite{TS}. This implies that splitting coefficients satisfying Eq. \ref{eq:10} to at least fourth order also satisfy Eq. \ref{eq:2} to at least fourth order. Compositions with more stages and higher order SHO criteria can be exploited for significantly reducing higher order error terms. In this sense, the higher order SHO criteria are used only as a constraint for obtaining efficient splitting coefficients for more general Hamiltonian systems.\\

Analytical evaluation of splitting coefficients for more than three stages quickly becomes hopeless. Instead a numerical routine has been developed to satisfy the multi-objective optimization problem with variables $c$ and $d$ and minimization objective $\kappa$:
\begin{equation} \label{eq:17}
\begin{array}{l}
{\kappa _\lambda } \equiv \left| {{q^{\left( \lambda  \right)}}\left( {{t_i}} \right) - \lambda !{\zeta _\lambda }\left( {c,d} \right)} \right| = 0,\lambda  = 1,2,...,{\lambda _H}\
\end{array}\
\end{equation}
where $\lambda_H$ is referred to as the SHO order constraint and $\kappa_{\lambda}$ for $\lambda=1,2,...,\lambda_H$ are the minimization objectives. Numerical experiments suggest that for $\lambda_H>5$ every additional stage yields two increments of $\lambda_{max}$ and a single increment of $\lambda_H$ for which at least one real solution exists.\\

The \textit{BAB} decomposition is numerically evaluated with the choice of constants in Eq. \ref{eq:12}: $a=-b=1$ and $m=k=1$ corresponding to the initial conditions $q_a=-p_a=1$. The objectives $\kappa_0$ and $\kappa_1$ are always 0 if $\sum\nolimits_{i = 1}^k {{c_i}}  = \sum\nolimits_{i = 1}^k {{d_i}}  = 1$. Additionally these identities eliminate one $c$ and $d$ variable from the minimization procedure. The minimization is performed using an adaptive simplex method \cite{GA} minimizing only one objective $\kappa_{max}=max(\kappa_1,\kappa_2,...,\kappa_{\lambda} )$. Solutions are provided in the following section with minimization criteria $\kappa_{max}<1e-128$ calculated using software-implemented arbitrary precision arithmetic (2048bit, 256bit exponent). Four unique sets of splitting coefficients are presented in the following section. The non-unique solutions are picked among thousands of solutions based on the sum of the absolute values of the splitting coefficients and higher order SHO errors.\\\\

\section{Results}
\subsection{Obtained splitting coefficients}
Splitting coefficients with five to nine stages are presented in Table \ref{table:4} in array format for convenient implementation for various programming languages.\\

\noindent Two different kinds of \textit{BAB} solutions are evaluated:
\begin{itemize}\itemsep6pt \parskip0pt \parsep0pt
\item \textbf{BAB}: Solution to Eq. \ref{eq:10} using \textit{BAB} decomposition (see Table \ref{table:2}).
\item \textbf{BAB'}: Solution to Eq. \ref{eq:10} using \textit{ABA} decomposition with coefficients $c_0=c_{s+1}=0$ in Eq. \ref{eq:3}.
\end{itemize}
The difference between the \textbf{BAB} and \textbf{BAB'} solutions is revealed by Eq. \ref{eq:5} where the error of $p_{s+1}$ remains unaddressed for the \textbf{BAB} solution as opposed to \textbf{BAB'} solution where the error of $q_{s+1}$ remains unaddressed. Thus the \textbf{BAB} and \textbf{BAB'} solutions possess different constraints which may influence their performance in either direction.\\

\textbf{BAB} and \textbf{BAB'} solutions should be implemented according to Eq. \ref{eq:4}. Unique solutions satisfy both the \textit{ABA} and \textit{BAB} decompositions and can therefore be implemented as either, although the performance of the presented solutions depends on the applied scheme.\\

The naming convention of the obtained methods is [\textit{scheme}]s[\textit{stages}]o[\textit{order}]H, for instance \textbf{BAB's9o7H} for a \textit{BAB}-optimized scheme with nine stages and seventh harmonic order.\pagebreak
\nopagebreak

\begin{table*}[t] 
\begin{center}
\caption{Obtained splitting coefficients with 77 digits. The methods are organized according to their stage count.}
\label{table:4}
\centering
\begin{tabular}{l}

\bottomrule
Unique \textbf{ABAs5o6H}, \textbf{A}, \textbf{B} and \textbf{C}\\
\toprule

\scriptsize{\pbrow{
d(1) = 0.1558593591762168313166117535752091422239663993391011462498104831549442591694\\
d(2) =-0.0070254990919573173514483364758218294773716640092220571342056284758867609611\\
c(1) =-0.6859195549562166768601873150414759494319985863677163820719179393682014399373\\
c(2) = 0.9966295909529363159571451429325843698583459772292551181721475637244006507927\\\\
d(1) = 0.4020196038964999834667409950496227775945673320979099323902806525851620445492\\
d(2) = 0.5329396856308538150258772262086702929451721575835842834460326556965220312130\\
c(1) = 0.9110842375676615218574607388486783304139753525628699898390474132061253024968\\
c(2) = 0.1740059542332660799009374186088931171982348451547482386207462271424421679090\\\\
d(1) = 0.1868565631155112597511173758337610451623768791420295598869906080256347098408\\
d(2) = 0.5520581660514781484261043096825685955052553493857487316732455515112095793516\\
c(1) = 0.5642486163110637621453746447826190031465518453448216443978248524452914255263\\
c(2) =-0.2393627021773294286793711975145735718917010075899623225091609656425715483488
}}\\
\midrule
\scriptsize{\pbrow{
d(3)=0.5-d(1)-d(2);d(4)=d(3);d(5)=d(2);d(6)=d(1);\\
c(3)=1-2*(c(1)+c(2));c(4)=c(2);c(5)=c(1);
}}\\

\bottomrule
Unique \textbf{BABs6o7H} (more exist), \textbf{BABs6o5H} and \textbf{BAB's6o5H} \\
\toprule

\scriptsize{\pbrow{
d(1) = 0.0832701092493097690276300822599156817795619881080575430174826369500044839553\\
d(2) = 0.3997273690963360211284395920007795550575060531634793748020207288976344439468\\
d(3) =-0.0541842778124726964199287659702152862181671805554302053695494244408226729818\\
c(1) = 0.2475471587650765967910125296669232190787926795528258860075742877866898482465\\
c(2) = 0.5446579217808193419580029125986805136192611468678745304306198457355253495088\\\\
d(1) = 0.0658831533161155021794371297629949214211270641450367882108652454225238292357\\
d(2) =-0.6711629060948253965117521242801468651670183829736696004743743104248379033034\\
d(3) = 0.9736703100725350498414312651550857191131218932308788320806762063004096320895\\
c(1) = 0.2265023974336291596186923088995152371194987043433278784212774210853229477086\\
c(2) =-0.0047799986678794678665602622568725658855054645768977416258774175978957628102\\\\
d(1) = 0.0650508268637574949487516678539036744380576078003520231297474895927182047842\\
d(2) =-0.3948051939117155639582651907195511796839512131373933326629623631591978696178\\
d(3) = 0.6918498547904058960782554213200966000604457266088855079657967408955821538731\\
c(1) = 0.2328962665845291347812910553597276545034489573682700034501734659308763580659\\
c(2) =-0.0111617638003721094728940473306267483522869816097800738075975236192041340802
}}\\
\midrule
\scriptsize{\pbrow{
d(4)=1-2*(d(1)+d(2)+d(3));d(5)=d(3);d(6)=d(2);d(7)=d(1);\\ 
c(3)=0.5-c(1)-c(2);c(4)=c(3);c(5)=c(2);c(6)=c(1);
}}\\
\bottomrule
\textbf{BABs7o7H} and \textbf{BAB's7o6H}\\
\toprule
\scriptsize{\pbrow{
d(1) = 0.0638745574250616045658401356462756092272737349204789877616691621039130680037\\
d(2) =-0.0650239777505938311516598494765811300128929849501107531440553736739769298894\\
d(3) = 0.2509446105745547370613575645855473357282136355718617210088090794709222342775\\
c(1) = 0.2752781729059777393394978710448690782125215018949186085075325605348526197756\\
c(2) =-0.0843138705589167473554015820986490036832890668438279781819362930106920807542\\
c(3) = 0.1674497222006475614401177016323447087805836086414469568091358611098423440220\\\\
d(1) = 0.0522155297747848201407012160969040693245471580104797248381281194964273517726\\
d(2) =-0.0824972558529561412131911937717420514162728339681056503508469680313691406287\\
d(3) = 0.3285541797987193353601113204079269672646845923663727576986276602617960257026\\
c(1) = 0.2487563308365098625528031803769571289196558939258433219240690943143969198069\\
c(2) =-0.0651011247076581799932061212576878177123945470202647856511921757791017775052\\
c(3) = 0.2480624780675545152650672751613106579864581926645260137078906816928505888862
}}\\
\midrule
\scriptsize{\pbrow{
d(4)=0.5-d(1)-d(2)-d(3);d(5)=d(4);d(6)=d(3);d(7)=d(2);d(8)=d(1);\\
c(4)=1-2*(c(1)+c(2)+c(3));c(5)=c(3);c(6)=c(2);c(7)=c(1);
}}\\
\bottomrule
\textbf{BAB's8o7H}\\
\toprule
\scriptsize{\pbrow{
d(1) = 0.0538184115480034769403763798524605188562842390760879592632218376015166638395\\
d(2) = 0.1648743326910472361014809085317059425299121141031052090901977952513984878990\\
d(3) = 0.3895399407808198068744134256203146340834631254960864069726823667050364522355\\
d(4) =-0.2288957415563594299572505173565338312542463595622272333825061768110435645417\\
c(1) = 0.1486140577445185629163082471176700173109512976367237631150576219945233462284\\
c(2) = 0.1071986675806227950500566279939336794589433458464489776124879870581484936262\\
c(3) =-0.0149646736494517061945681450558142918831874360003431672632178480031079700216
}}\\
\midrule
\scriptsize{\pbrow{
d(5)=1-2*(d(1)+d(2)+d(3)+d(4));d(6)=d(4);d(7)=d(3);d(8)=d(2);d(9)=d(1);\\
c(4)=0.5-c(1)-c(2)-c(3);c(5)=c(4);c(6)=c(3);c(7)=c(2);c(8)=c(1);
}}\\
\bottomrule
\textbf{BAB's9o7H}\\
\toprule
\scriptsize{\pbrow{
d(1) = 0.0464929004396589154281717058427105561306160230440930588914036807441235817244\\
d(2) = 0.1549010127028879927850680477816652638346460615901974901213193690401204696252\\
d(3) = 0.3197054828735917137611074311771339117602994884245091220333340037841616085048\\
d(4) =-0.1929200088157132136865513532391282410293753210475133631464188500663304857888\\
c(1) = 0.1289555065927298176557065467802633438775379080212831185779306825670371511433\\
c(2) = 0.1090764298548827040268039227200943338187149719339317536310302288046641781422\\
c(3) =-0.0138860356804715144111581981849964201100030653749527555344377031679795959892\\
c(4) = 0.1837549745641803566768357217228586277331494085368674804908537743649129597425
}}\\
\midrule
\scriptsize{\pbrow{
d(5)=0.5-d(1)-d(2)-d(3)-d(4);d(6)=d(5);d(7)=d(4);d(8)=d(3);d(9)=d(2);d(10)=d(1);\\
c(5)=1-2*(c(1)+c(2)+c(3)+c(4));c(6)=c(4);c(7)=c(3);c(8)=c(2);c(9)=c(1);
}}\\
\bottomrule
\end{tabular}
\end{center}
\end{table*}

\subsection{Numerical tests}
Common methods used to benchmark SIs for integration of autonomous Hamiltonian systems include \cite{WI} \cite{BL} \cite{FA}:
\begin{itemize} \itemsep6pt \parskip0pt \parsep0pt
  \item The maximum Hamiltonian error: $max(| \Delta H(t)/{H_0} |)$
  \item The mean Hamiltonian error: $mean(| \Delta H(t)/{H_0} |)$
  \item The accumulated error of integrals
\end{itemize}
where $H_0$ is the initial Hamiltonian. Each of these benchmark methods are applied throughout this section.\\

\noindent The following 10 symmetric SIs are benchmarked for three autonomous Hamiltonian systems: The simple harmonic oscillator, the Hénon Heiles system and the Kepler problem:
\begin{itemize} \itemsep0pt \parskip0pt \parsep0pt
\item \textbf{Ruth [ABA]} 3-stage, 4th order \cite{RU}
\item \textbf{s5odr4 [ABA]} 5-stage, 4th order \cite{KA}
\item \textbf{ABA104 [ABA]} 7-stage, 10-4 generalized order \cite{BL}
\item \textbf{ABA864 [ABA]} 7-stage, 8-6-4 generalized order \cite{BL}
\item \textbf{ABA1064 [ABA]} 8-stage, 10-6-4 generalized order \cite{BL}
\item \textbf{Yosh s7o6 A [ABA]} 7-stage, 6th order \cite{YO}
\item \textbf{ABAs5o6H [ABA]} 5-stage, 4th order
\item \textbf{BABs7o7H [BAB]} 7-stage, 4th order
\item \textbf{BAB's8o7H [BAB]} 7-stage, 4th order
\item \textbf{BAB's9o7H [BAB]} 9-stage, 4th order
\end{itemize}
where \textbf{[ABA]} and \textbf{[BAB]} denote suggested \textit{ABA}- and \textit{BAB}-implementation respectively. It should be noted that the generalized order schemes are designed for perturbed Hamiltonian systems $H = H_{A}+\epsilon H_{B}$ with $\epsilon \ll 1$. It should also be noted \textbf{Yosh s7o6 A} yields the smallest errors of the three stage-optimized sixth order integrators presented by \cite{YO}, and that this integrator is here evaluated to reference higher order accuracy and stability.\\

Numerical tests are performed using double precision floating point format with compensated summation. Compensated summation drastically reduces truncation errors for long-term symplectic integration (see \cite{KA} and references therein). All numerical integrations are performed using \textbf{Zymplectic} (v.1.01.00) from \href{http://zymplectic.com}{\textit{Zymplectic.com}}, which is a numerical framework designed for numerical integration and benchmark of separable Hamiltonian systems.\\

\begin{figure}[H]
\centering
\includegraphics[clip,scale=0.9]{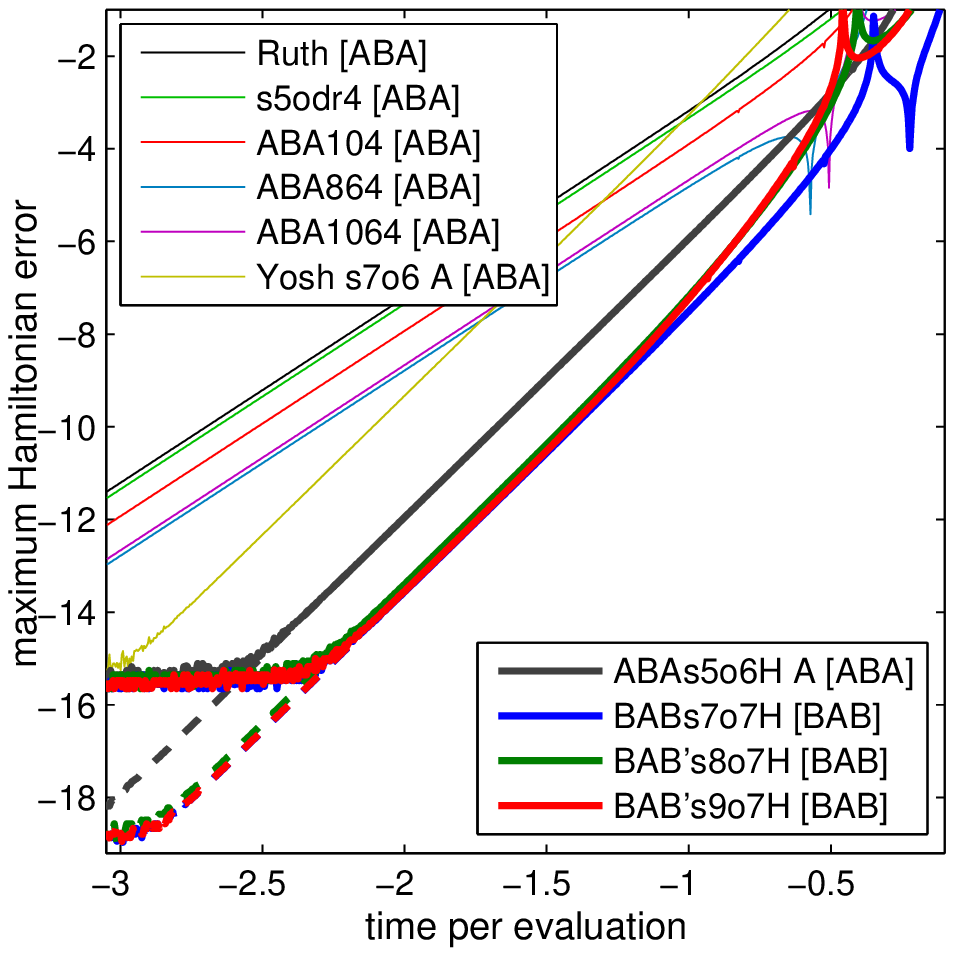}\hfill \break
\includegraphics[clip,scale=0.9]{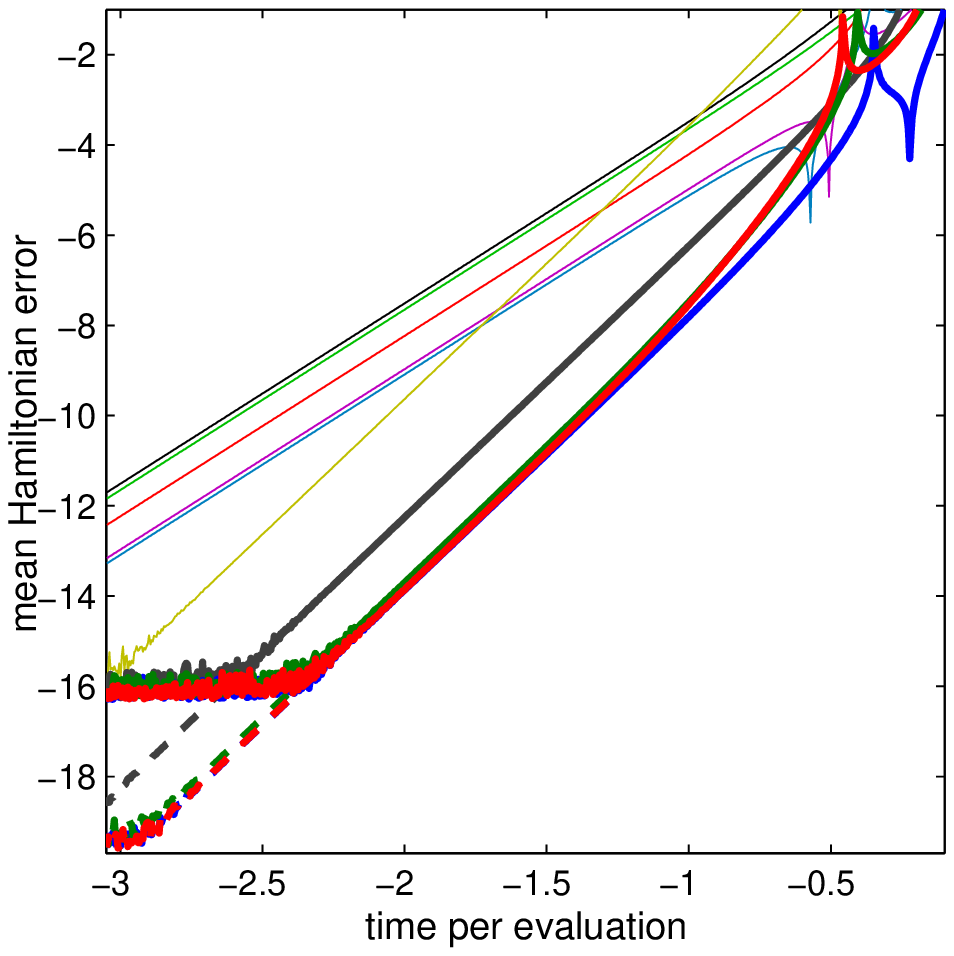}
\caption{Integrator benchmark using the SHO in the time interval $t \in [0,500]$. Top: Maximum Hamiltonian error. Bottom: Average relative Hamiltonian error. The dashed lines show benchmark profiles for integrators of corresponding color using extended precision. The horizontal axes correspond to $\tau/s$. All axes are in base 10 logarithmic units. Legend applies to both graphs}
\label{fig:1}
\end{figure}

\textbf{The simple harmonic oscillator}\\
Consider the Hamiltonian of the one-dimensional SHO (Eq. \ref{eq:11}) with $k=m=1$ and initial conditions $(p,q)=(0,1)$. Figure \ref{fig:1} shows a benchmark profile of the 10 integrators for the SHO. Every point on each line corresponds to a complete integration in a given time interval. The slopes of the lines read the integrator order. It is noticable that the integrators \textbf{ABAs5o6H A}, \textbf{BABs7o7H}, \textbf{BAB's8o7H} and \textbf{BAB's9o7H} behave as sixth order integrators.\\

The \textit{ABA}- and \textit{BAB}-decomposition from the previous section can be utilized to determine error contributions of different orders. More precisely, $\kappa_{\lambda}$ can be evaluated for all $\lambda$ to resolve the magnitude of higher order errors for the SHO. This can be used to characterize higher order error terms which often diverge rapidly for high order stage-optimized integrators. While these error terms strictly apply to the SHO, they still cover  general equations present in the expansion of Eq. \ref{eq:2}.\\
\begin{figure}[H]
\centering
\includegraphics[scale=0.9]{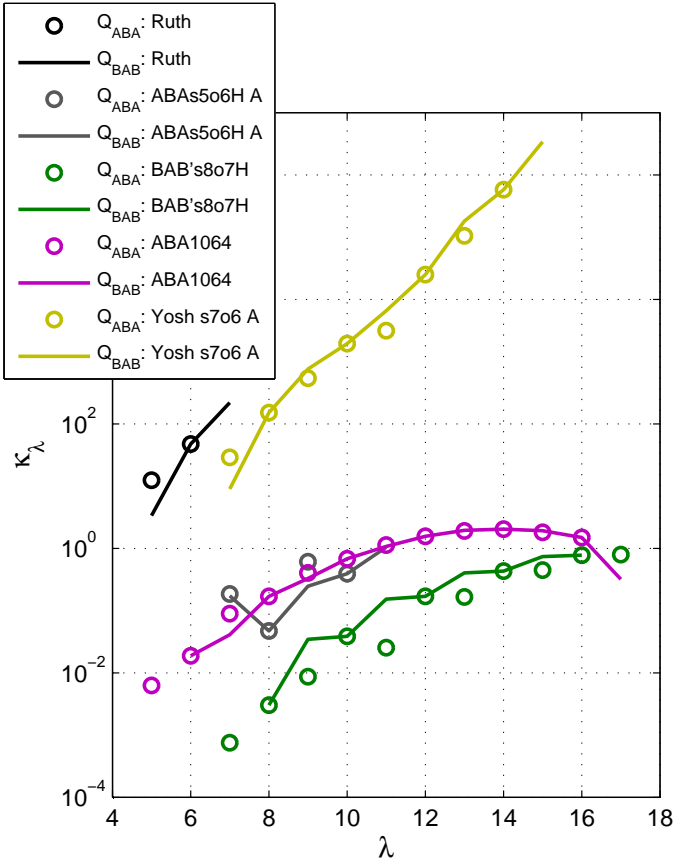}
\caption{SHO decomposition of selected integrators for both the \textit{ABA}- (circle marker) and \textit{BAB}-composition (solid line). Lines and markers of the same color intersect exactly at even orders}
\label{fig:2}
\end{figure}
\noindent Figure \ref{fig:2} shows position error terms  of selected integrators for the SHO. Notice that the obtained \textit{BAB} methods have more error terms when implemented as \textit{ABA} methods. Furthermore the seventh harmonic order of \textbf{BAB's8o7H} eliminates the seventh order error term of the position coordinate for the \textit{BAB} implementation as opposed to the sixth harmonic order integrator \textbf{ABAs5o6H A}. The largest integer value of $\lambda$ with no integration error reads the SHO integration order. Stage-optimized integrators of higher orders generally have rapidly diverging higher order error contributions. This implies that these integrators are not suitable for integrating the SHO with large time steps.\\ 

\begin{figure}[H]
\centering
\includegraphics[clip,scale=0.9]{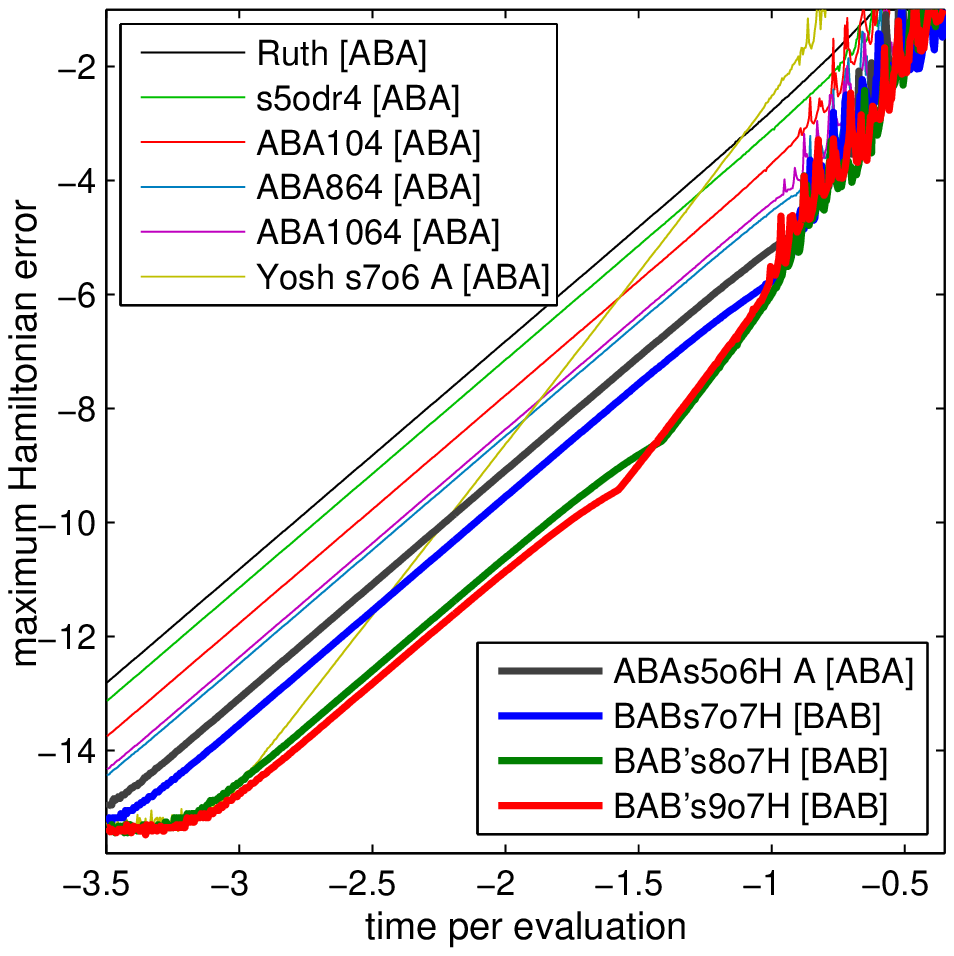}\hfill \break
\includegraphics[clip,scale=0.9]{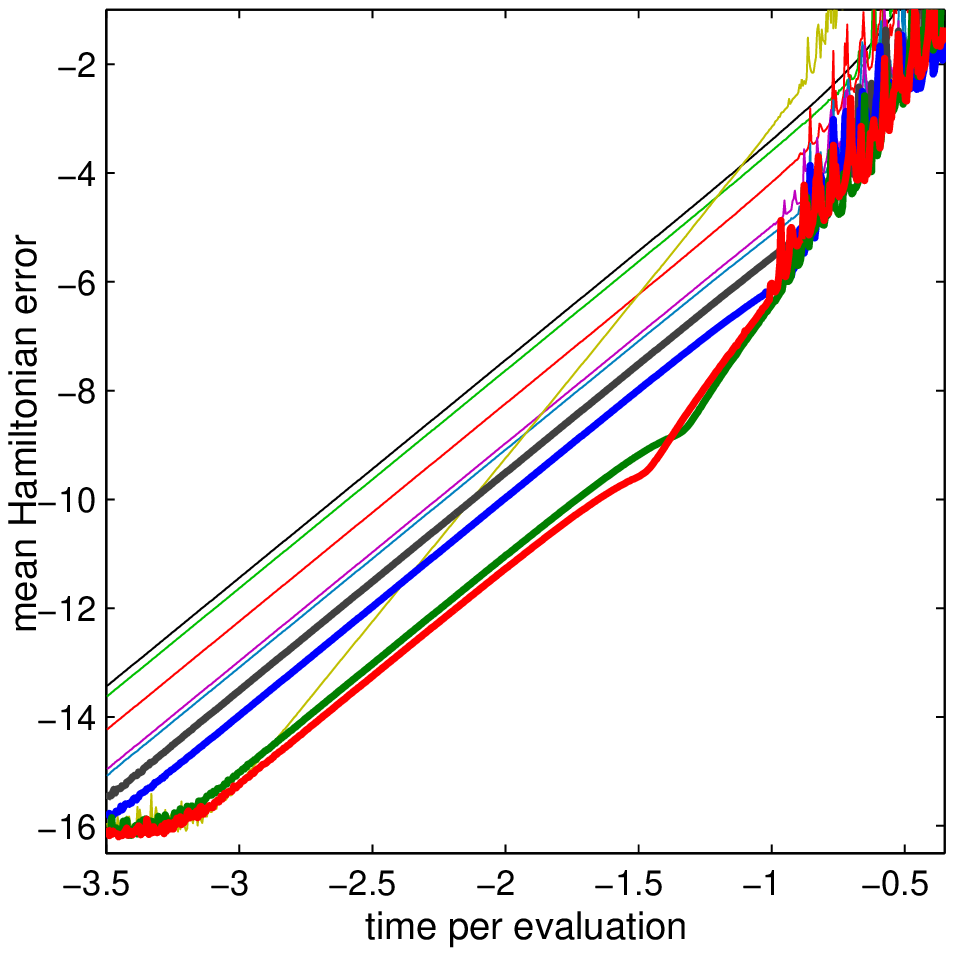}
\caption{Integrator benchmark using the Hénon Heiles system in the time interval $t \in [0,500]$. Top: Maximum Hamiltonian error. Bottom: Average relative Hamiltonian error. The horizontal axes correspond to $\tau/s$. All axes are in base 10 logarithmic units. Legend applies to both graphs}
\label{fig:3}
\end{figure}

\textbf{The Hénon Heiles system}\\
The Hénon-Heiles system is one of the most studied Hamiltonian systems due to its applications in celestial mechanics, its non-integrable properties, chaotic dynamics and fractal structures. Consider the Hamiltonian of the Hénon-Heiles system with initial conditions $[q_x,q_y,p_x,p_y]=[0.3,0.0,0.0,0.4]$:
\begin{equation} \label{eq:18}
\begin{array}{l}
H = \frac{{p_x^2 + p_y^2}}{2} + \frac{{{q_x}^2 + {q_y}^2}}{2} + {q_x}^2{q_y} - \frac{1}{3}{q_y}^3\
\end{array}\
\end{equation}
where in this case $H=1/8$, inside the chaotic domain. Figure \ref{fig:3} shows a benchmark profile of the 10 integrators for the Hénon Heiles system. The Hénon Heiles system can be considered as a perturbed SHO for small values of $q_x$ and $q_y$. For this reason it appears that especially \textbf{BAB's8o7H} and \textbf{BAB's9o7H} exhibit a sixth order descent of error for large $\tau$. 

\textbf{The Kepler problem}\\ 
Symplectic integration has been widely used for high accuracy simulations of the Solar System (see \cite{FA} and references therein). It has been proposed that Mercury due to its fast orbital period and high eccentricity is limiting the performance of SIs when integrating the Solar System \cite{FA}. Consider the Hamiltonian of the planar Sun-Mercury system: 
\begin{equation} \label{eq:19}
\begin{array}{l}
H = \frac{{{{\left\| \mathbf{p} \right\|}^2}}}{{2M}} - G\frac{{{m_S}{m_M}}}{{\left\| \mathbf{q} \right\|}} \approx \frac{{{{\left\| \mathbf{p} \right\|}^2}}}{{2{m_M}}} - G\frac{{{m_S}{m_M}}}{{\left\| \mathbf{q} \right\|}}\
\end{array}\
\end{equation}
where $M$ is the reduced mass, $G$ is the gravitational constant, and $m_S$ and $m_M$ are the masses of the Sun and Mercury respectively. The planar Sun-Mercury system is here evaluated using data from \href{http://nssdc.gsfc.nasa.gov/}{\textit{nssdc.gsfc.nasa.gov}} where the aphelion is chosen as the initial Sun-Mercury distance yielding the minimum orbital velocity.\\

\begin{figure}[H]
\centering
\includegraphics[clip,scale=0.9]{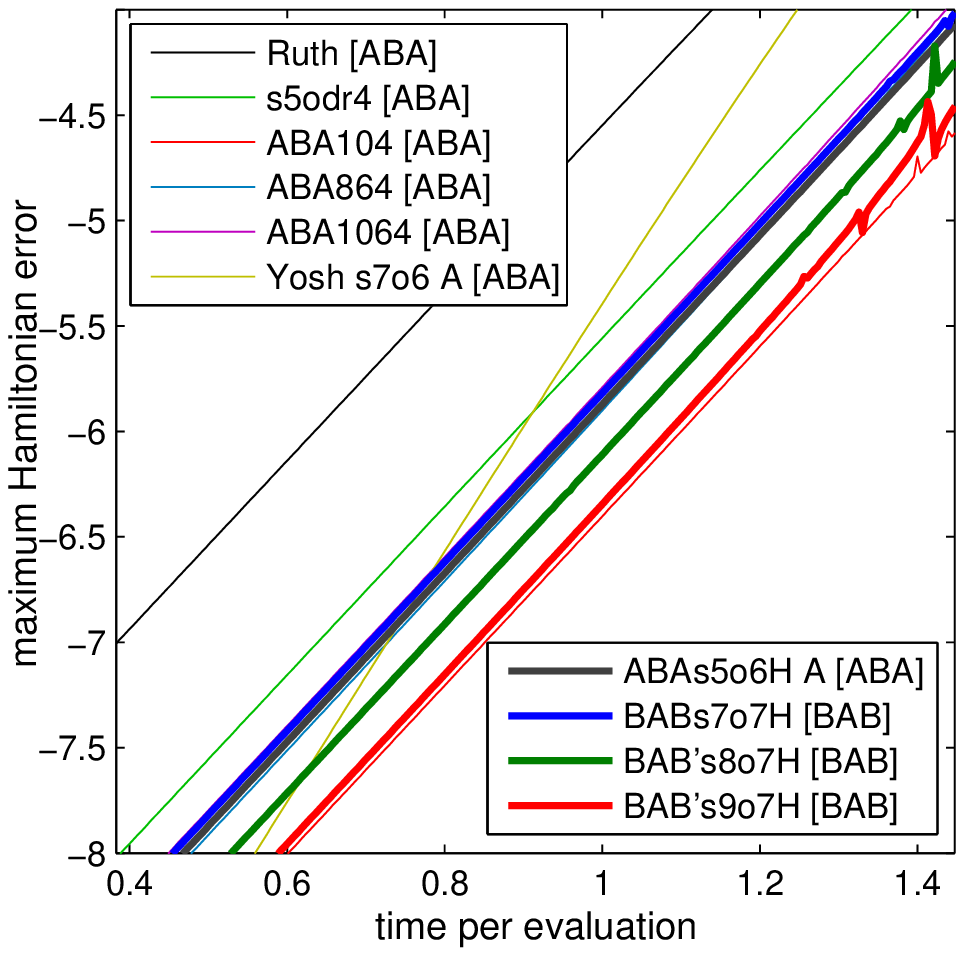}\hfill \break
\includegraphics[clip,scale=0.9]{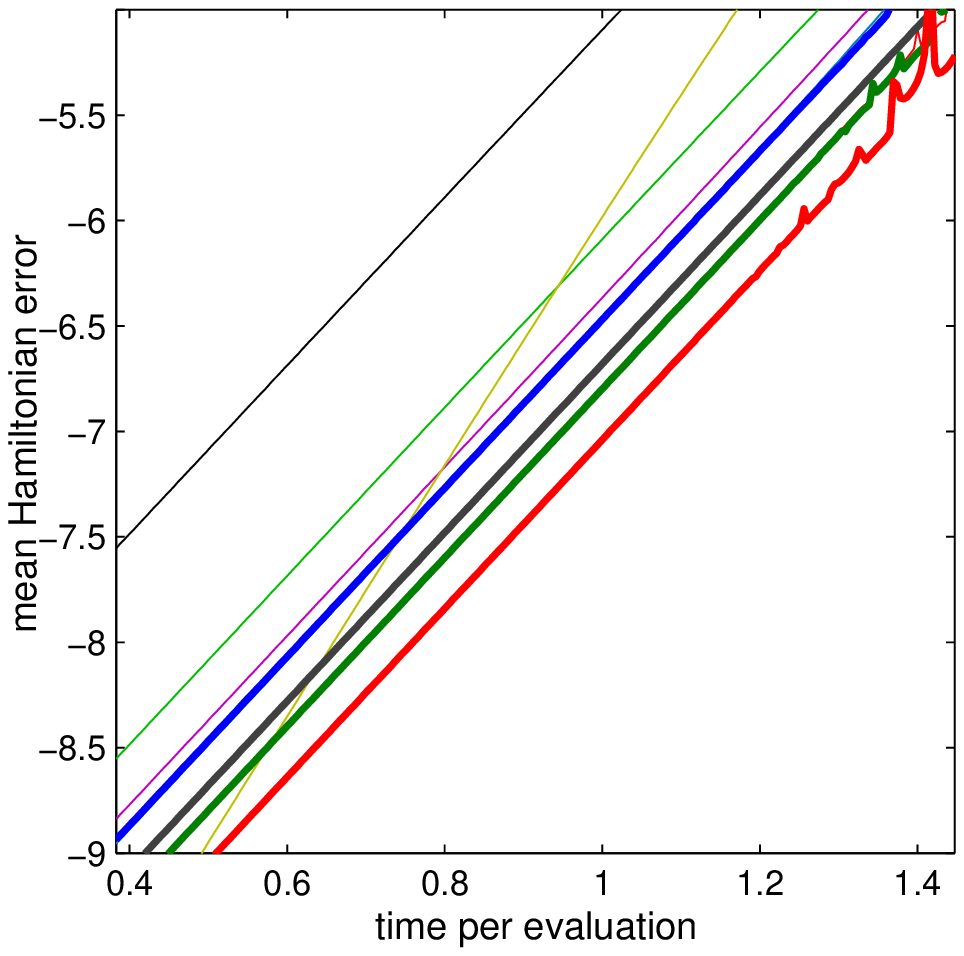}
\caption{Integrator benchmark using the planar Sun-Mercury system in the time interval $t \in [0,10yr]$. Top: Maximum Hamiltonian error. Bottom: Average relative Hamiltonian error. The horizontal axes correspond to $\tau/s$ in units of hours. All axes are in base 10 logarithmic units. Legend applies to both graphs}
\label{fig:5}
\end{figure}

Figure \ref{fig:5} shows a benchmark profile of the 10 integrators for the Sun-Mercury system. It is noticable that \textbf{ABA104} shows the best performance in regard to the maximum error and \textbf{BAB's9o7H} shows the best performance in regard to the average error.\\

\noindent The accumulated error in a two body system (Sun-Mercury) manifests as a secular advance of the orbital ellipse similar to the precession of perihelion induced by general relativity or the gravitational pull of other planets. The mathematical nature of the secular SI error is described by \cite{CH2}. The secular advance of the orbital ellipse can be evaluated through the rotation of the Laplace-Runge-Lenz vector $\mathbf{A}$:
\begin{equation} \label{eq:20}
\begin{array}{l}
\mathbf{A} = \mathbf{p} \times \mathbf{L} - \mathbf{\hat r}\
\end{array}\
\end{equation}
where $\mathbf{\hat r}=\mathbf{r}/r$, $\mathbf{L} = \mathbf{r} \times \mathbf{p}$, $\mathbf{p}$ is the momentum vector and $\mathbf{r}$ is the position vector with length $r$.
\begin{figure}[H]
\centering
\includegraphics[scale=0.9]{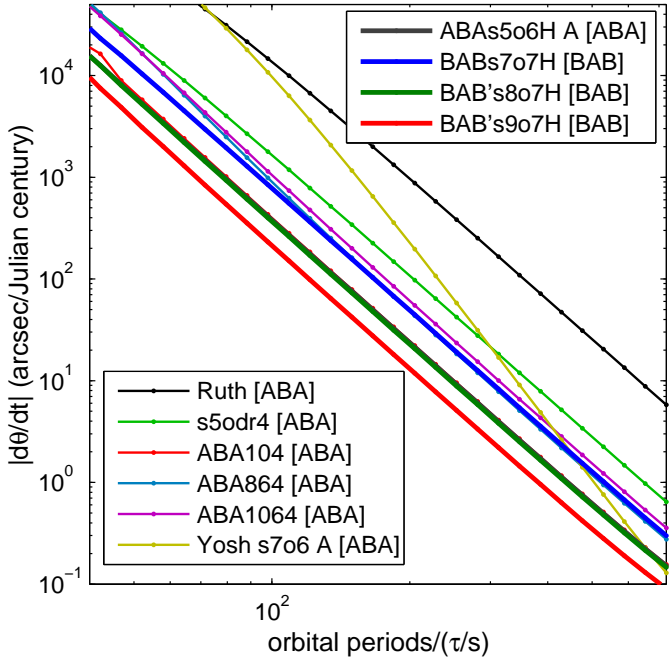}
\caption{Advance of perihelion (absolute rotation velocity) as a function of the number of evaluations per orbit. Linear fit $95\%$ confidence intervals of the perihelion advance are significantly smaller than the dot size. The integrators \textbf{ABAs5o6H}, \textbf{ABA104} and \textbf{BAB's8o7H} achieve approximately the same performance, obfuscating their plots}
\label{fig:4}
\end{figure}
\noindent The orbital precession advances linearly with time $t$ \cite{CH2}. The rotation $\theta$ is introduced as the angle of $\mathbf{A}$ implying that $\theta (t)$ is a linear function and is chosen such that $\theta (0)=0$. Figure \ref{fig:5} shows the perihelion precession $d \theta / dt$ which has been obtained numerically through linear regression of $\theta(t)$ over numerous orbital periods. Comparison of Figure \ref{fig:4} and \ref{fig:5} suggests that the accumulated error is in better agreement with the average error than the maximum error.
\end{multicols}
\begin{figure*}[t]
\begin{center}
\includegraphics[scale=0.80]{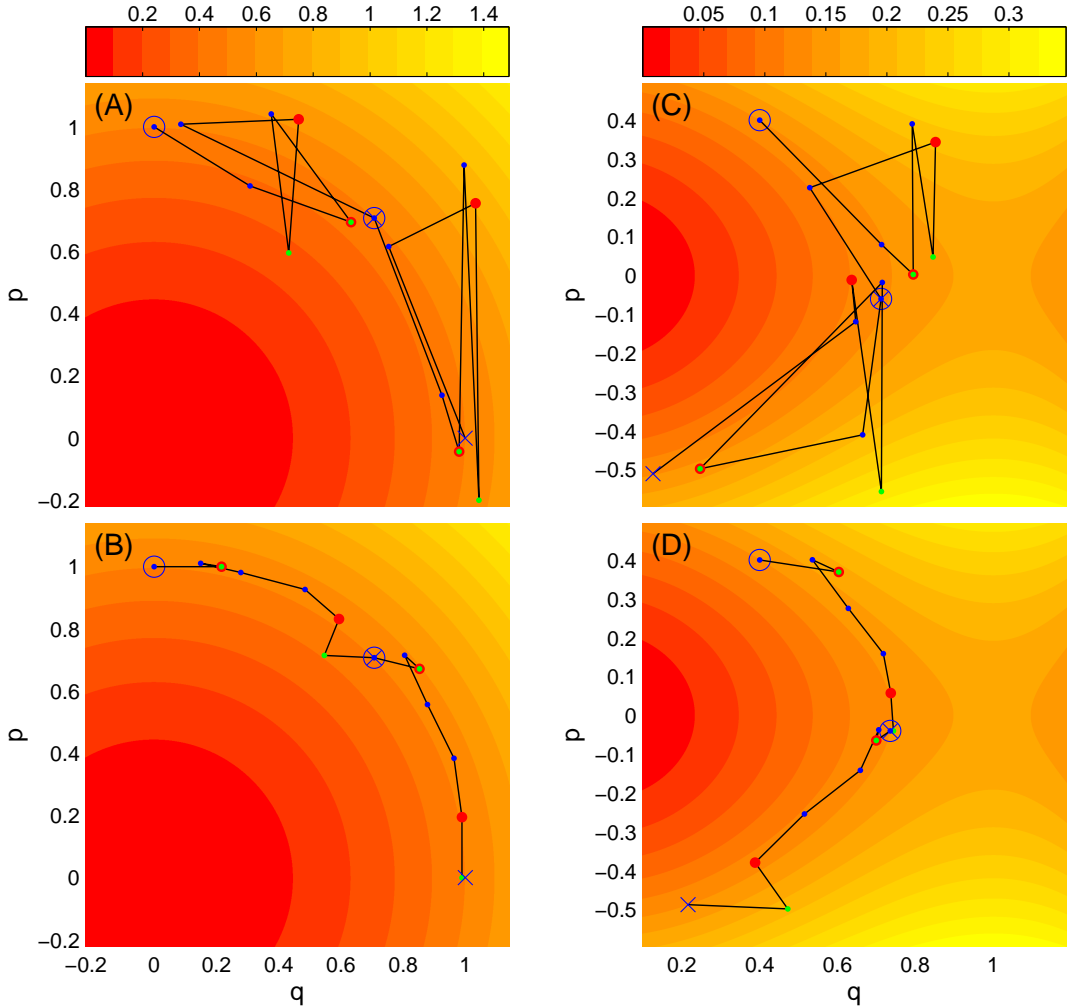}
\caption{Phase space illustration of the substeps performed over two iterations by \textbf{Yosh s7o6 A} (A)(C) and \textbf{BABs7o7H} (B)(D). The integrators are applied to the SHO (A)(B) with initial conditions $(q,p)=(0,1)$ and time step $\tau=\pi/4$, and the Hénon-Heiles system (C)(D) in the y-plane $(q_x=p_x=0)$ with initial conditions $(q,p)=(q_y,p_y )=(0.4,0.4)$ and time step $\tau=2$. Blue circles and Xs indicate the initial and final positions respectively for each iteration. Blue dots indicate substeps with positive $c_i$ and $d_i$ coefficient for calculating the next substep. Red dots and green dots indicate substeps with negative $c_i$ and $d_i$ coefficients respectively for calculating the next substep. A green dot within a red dot indicates that both $c_i$ and $d_i$ are negative. The contour colors display the Hamiltonian with values given by the colorbars. (C) shows an unstable integration where $H$ is not conserved. (A) (B) and (D) show stable integrations which preserve $H$ indefinitely}
\label{fig:6}
\end{center}
\end{figure*}
\begin{multicols}{2}
\section{Discussion}
A thorough benchmark analysis of 10 different integrators has been performed for the SHO, the Hénon Heiles system and the Kepler problem. All numerical tests suggest that the obtained integrators in spite of an increased number of stages greatly and consistently improve the integration accuracy and stability. In fact, the new schemes outperforms the acclaimed stage-optimized integrator \textbf{Ruth} by several orders of magnitude. The higher order behavior for SHO-perturbed systems is an additional perk of the obtained integrators. This property may be particularly useful in fields of molecular dynamics where symplectic correctors may not be suitable. Optimized fourth order schemes appear to be useful for achieving high numerical stability allowing large time steps with minimal integration error. High stability is particularly important for systems where the complexity of $H$ varies with time: Close encounters in the gravitational N-body system inevitably cause high errors ultimately waiving symplectic properties of SIs. Higher order integrators are generally more constrained and usually involve several negative splitting coefficients with large absolute values. The performance breakpoint of higher order integrators can be visualized by considering the executed substeps. Figure \ref{fig:6} illustrates the substeps performed by two seven stage SIs (\textbf{Yosh s7o6 A} and \textbf{BABs7o7H}) for the SHO and Hénon-Heiles system. Figure \ref{fig:6} reveals that \textbf{Yosh s7o6 A} (A) (C) performs large forward and backward steps compared to \textbf{BABs7o7H} (B) (D). The steps are too large in (C) for the Hamiltonian to be preserved. \textbf{BABs7o7H} and other splitting coefficients obtained in this paper generally perform small near-forward steps. Note that some of the methods presented in Table \ref{table:4} only have small coefficients for either $d$ or $c$. For this reason the integrator performance may change drastically if the roles of $T$ and $V$ are switched. The performance of the obtained integrators generally declines for systems where $T(p)$ deviates significantly from a squared kinetic energy. 

\section{Conclusion}
Efficient fourth order splitting coefficients have been obtained by satisfying higher order conditions for the simple harmonic oscillator. Numerical tests indicate that the obtained integrators outperform the three stage fourth order integrator by R. Ruth as well as the more optimized integrators by W. Kahan and S. Blanes. The obtained integrators also compare favorably with higher order integrators by H. Yoshida for moderate time steps. The integrators are particularly suitable for particle systems that require high numerical stability and accuracy at large time steps. The best performance is generally achieved by the methods \textbf{BAB's9o7H}, \textbf{BAB's8o7H} and \textbf{ABAs5o6H A}.\\

It has been found that harmonic order conditions higher than seven often introduce more negative splitting coefficients. A larger number of stages presents more degrees of freedom resulting in more inappropriate solutions. Additional constraints should be pursued to obtain efficient splitting coefficients with more than nine stages.\\
\indent The presented decomposition can be used to satisfy general fourth order criteria and higher order conditions specifically for the simple harmonic oscillator. This work suggests that general purpose high accuracy, high stability explicit symplectic integrators can be pursued by approaching multiple higher order conditions. This has been achieved here by satisfying individual high order entries.

\end{multicols}
\center\rule{10cm}{0.4pt}
\begin{multicols}{2}

\end{multicols}
\end{document}